\documentclass[11pt,a4]{article} 
\usepackage{makeidx}

\newif\ifpdf
\ifx\pdfoutput\undefined
   \pdffalse
\else
   \pdfoutput=1
   \pdftrue
\fi

\ifpdf
  \usepackage[
  pdftex,
  hyperindex,
  pagebackref,
  colorlinks,
  linkcolor=blue,
  menucolor=blue,
  pagecolor=blue,
  urlcolor=blue,
  citecolor=blue,
  pdftitle={A Note on the GNSS-R Interferometric Complex Field Coherence Time},
  pdfauthor={Giulio Ruffini},
  pdfcreator={LaTeX}
  ]{hyperref}

  \usepackage{thumbpdf}
  \usepackage{url}
  \usepackage[pdftex]{graphicx}
  \DeclareGraphicsExtensions{.pdf,.jpg,.mps,.png} 

\else

  \usepackage[dvips]{graphicx}
  \DeclareGraphicsExtensions{.eps,.jpg,.mps,.png,.ps}

\fi                          

\usepackage{fancyhdr}
\usepackage{multicol}     
\graphicspath{{figures/}}
\usepackage[english]{babel}
\usepackage[]{times}

\catcode`\@=11

\@addtoreset{equation}{section}
\def\theequation{\arabic{section}.\arabic{equation}}

\catcode`\@=11

\def\section{\@startsection{section}{1}{\z@}{3.5ex plus 1ex minus
   .2ex}{2.3ex plus .2ex}{\large\bf}}

\def\eqnarray{\let\@currentlabel=\theequation\refstepcounter{equation}
    \global\@eqnswtrue
    \global\@eqcnt\z@\tabskip\@centering\let\\=\@eqncr
    $$\halign to \displaywidth\bgroup\@eqnsel\hskip\@centering
      $\displaystyle\tabskip\z@{##}$&\global\@eqcnt\@ne
       \hfil${{}##{}}$\hfil
      &\global\@eqcnt\tw@ $\displaystyle\tabskip\z@{##}$\hfil
       \tabskip\@centering&\llap{##}\tabskip\z@\cr}
\def\lefteqn#1{\hbox to 4\arraycolsep{$\displaystyle #1$\hss}}
\def\thesection{\arabic{section}.}

\def\appendix{\setcounter{section}{0}
        \def\thesection{Appendix.}
        \def\theequation{\Alph{section}.\arabic{equation}}}

\long\def\@makefntext#1{\parindent 0cm\noindent
\hbox to 1em{\hss$^{\@thefnmark}$}#1}
\def\IR{{\hbox{{\rm I}\kern-.2em\hbox{\rm R}}}}
\def\IH{{\hbox{{\rm I}\kern-.2em\hbox{\rm H}}}}
\def\IC{{\ \hbox{{\rm I}\kern-.6em\hbox{\bf C}}}}
\def\IZ{{\hbox{{\rm Z}\kern-.4em\hbox{\rm Z}}}}

\newcommand{\beq}{\begin{equation}}
\newcommand{\be}{\begin{equation}}
\newcommand{\eeq}{\end{equation}}
\newcommand{\ee}{\end{equation}}
\newcommand{\bea}{\begin{eqnarray}}
\newcommand{\eea}{\end{eqnarray}}
\newcommand{\bean}{\begin{eqnarray*}}
\newcommand{\eean}{\end{eqnarray*}}
\newcommand{\ba}{\beq\begin{array}{lll} }
\newcommand{\ea}{\end{array}\eeq}

\def\IC{ {\rm l\hspace{-1.2ex}C} }    
\def\IZ{{\hbox{{\rm Z}\kern-.4em\hbox{\rm Z}}}}
\def\IR{{\hbox{{\rm I}\kern-.2em\hbox{\rm R}}}}
\newcommand\unmarkfootnote[1]{%
  \begingroup
    \let\@makefntext\noindent
    \footnotetext{#1}%
  \endgroup}

\newcommand{\captionfonts}{\footnotesize}

\makeatletter  
\long\def\@makecaption#1#2{%
  \vskip\abovecaptionskip
  \sbox\@tempboxa{{\captionfonts #1: #2}}%
  \ifdim \wd\@tempboxa >\hsize
    {\captionfonts #1: #2\par}
  \else
    \hbox to\hsize{\hfil\box\@tempboxa\hfil}%
  \fi
  \vskip\belowcaptionskip}
\makeatother   

\begin{document}

\thispagestyle{empty}

\begin{titlepage}
\vspace{.5in}

\vspace{2.5in}

\begin{center}

{\Large\bf On the GNSS-R Interferometric Complex Field Coherence Time}\\ \vspace{1.3in}


\begin{figure}[h!]
\centering
 \includegraphics[width=5cm,angle=0]{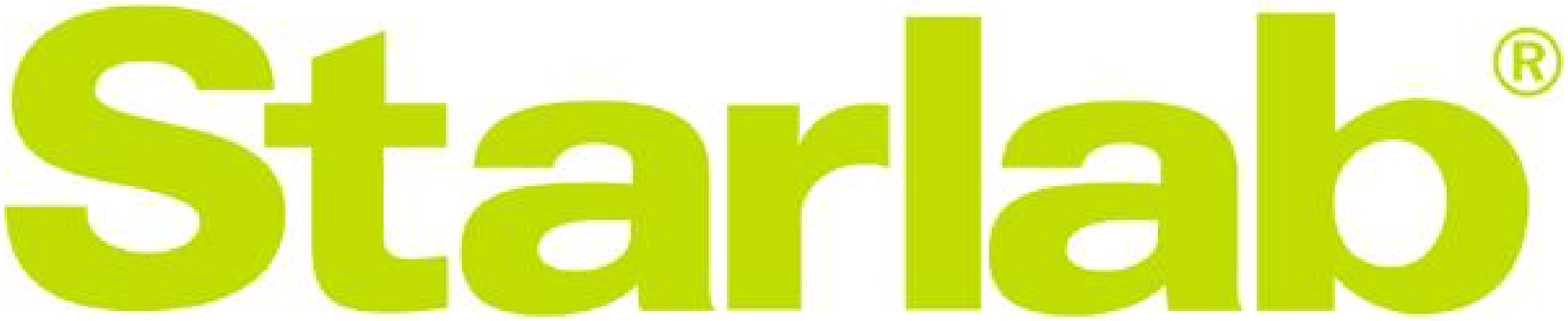} 
\end{figure}
{\large \bf Starlab Technical Brief TB0005}\\

\vspace{.4in} 

   {Giulio~Ruffini\footnote{\it email: giulio.ruffini@starlab.es} and Fran\c cois Soulat\\
        {\small\it Research Department, Starlab}\\
{\small\it Ed. de l'Observatori Fabra, C. de l'Observatori, s/n} \\
{ \small\it 08035 Barcelona, Spain \\ Compiled \today}
        } 
 \end{center}

 \vspace{.2in}
\begin{center}
{\large\bf Abstract}
\end{center}
\begin{center}
\begin{minipage}{5.4in}
{\small  In this paper we focus on the microwave bistatic scattering process, with the aim of deriving an expression for the interferometric complex field auto-correlation function from a static platform. We start from the Fresnel integral and derive the auto-correlation function in the Fraunhofer and Modified Fraunhofer regime. The autocorrelation function at short times can be expressed as a Gaussian with a direction dependent time scale. The directional modulation is a function of the angle between the scattering direction and the wave direction. The obtained relation can be used for directional sea state estimation using one or more GNSS-R coastal receivers. }
\medskip

    \end{minipage}
    \end{center} 
\end{titlepage}




\section{Introduction}

The use of Global Navigation Satellite System (GNSS) signals reflected by the sea surface as a remote sensing tool has generated considerable attention for over a decade.
Among several applications, two classes have rapidly emerged in the community:
sea-surface altimetry, which aims at retrieving the mean sea level like classical radar altimeters do, and sea-surface scatterometry or ``speculometry'' (a better term, as the GNSS-R return is dominated by quasi-specular scattering)
 for the determination of sea roughness and near surface wind. This paper addresses a new application: sea state monitoring from low altitude, static platforms.


The  Oceanpal project at Starlab focuses on the development of technologies for operational in-situ or low-altitude water surface monitoring  using GNSS-R. Oceanpal\footnote{Patent pending.} is a ``down-to-Earth'' offspring of technology developed within several ESA/ESTEC projects targeted on the exploitation of GNSS Reflections from space, following the proposal of   Mart\'\i n-Neira in 1993 \cite{martin-neira1993}. This instrument is to provide low cost/low maintenance sea state and altimetry measurements for coastal applications with the inherent precision of GNSS technology. Altimetric applications from the ground and from space have already been demonstrated in other campaigns \cite{caparrini2003,ruffini2004}, as well as airborne sea-state applications \cite{germain2003}. 

The present paper was motivated by a recent Oceanpal experiment conducted by Starlab in collaboration with the Barcelona Port Authority for the demonstration of GNSS-R sea state monitoring from low altitude, static platforms \cite{soulat2004}.  The HOPE 2 (Harbor Oceanpal Experiment) experimental campaign gathered GNSS-R  data from a ground platform in a range of sea-state conditions. It was carried out during December 2003 at the Meteorological station of Porta Coeli belonging to  the  Barcelona Port Authority, located on the port breakers. Two antennas were deployed at approximately  25 meters  over the open sea surface to gather GPS-L1 signals. As usual in GNSS-R, one antenna was zenith-looking (Right Hand Circularly Polarized)  to get the direct signal, and the other was oriented towards the open sea surface with the complementary polarization (Left Hand Circularly Polarization) to get the reflected signals. The  antennas were connected to the instrument Front End, which then generated  a digital bit-stream of unprocessed GPS  data at IF. The IF data were recorded at a sample frequency of $\sim$16 MHz, processed on board to produce complex waveforms, and transferred  to the instrument  mass storage unit.

Our analysis for sea state begins with the interferometric complex field (ICF), defined at time $t$ 
by $F_I(t) = F_R(t)/F_D(t) $,
where $F_D$ and $F_R$ are the complex values at the amplitude peaks of the direct and reflected complex waveforms, respectively. 
The direct signal is thus  used  as a reference to remove features unrelated to ocean motion, such as any residual Doppler, the navigation bit phase offset, or direct signal power variability. From a static platform, the ICF time history contains very valuable information on the sea state.

The goal of the present analysis is to relate dynamics of the ICF to sea-surface geophysical parameters. Given the static character of the experiment and the  very small scattering area---we recall the instrument  was deployed at low altitude ($\sim 25$ m) in this experiment---sea-surface roughness parameters such as the Directional Mean Square Slope (DMSS) are not foreseen to be quantifiable  through the estimation of the width and orientation of the scattering area (Delay-Doppler mapping) as they were in \cite{germain2003}---especially given the coarse nature of the C/A code which is available today.

In this note we formulate an expression for the ICF autocorrelation function in the Fraunhofer and modified Fraunhofer approximations starting from the Fresnel integral for the scattered field (scalar physical optics).

\section{Fresnel integral}
We start our derivation from the  Fresnel integral approximation for the scattered field and ignore for the time being curvature related polarization effects  (see p. 380, Eq. 17 in \cite{born}). The context is that of GPS reflections in a static receiver scenario. Although the GPS signal is not monochromatic, it can be thus approximated, since the code modulation (here represented by $a[t]$) adds relatively small bandwidth to the signal ($\omega_m << \omega$). This is an important consideration, since it means that there is no need to work with wave-packets. 

\begin{figure}[t!]
\centering
 \includegraphics[width=8.5cm,angle=270]{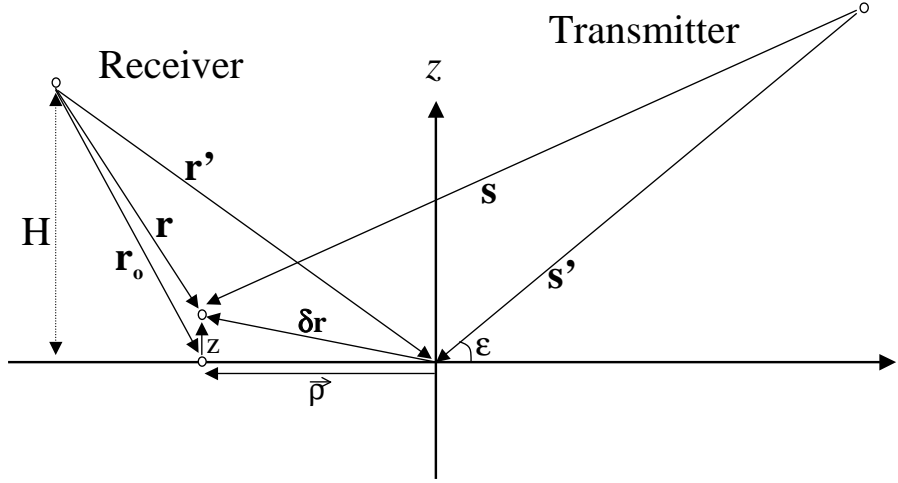} 
   \caption{Definition of vectors used in analysis. The specular point is located at the origin.}
   \label{geometry}
\end{figure}

We start from a signal of the form $a[t]\exp[i\omega t]\exp[in\pi]$, as emitted by the satellite (the last term is the navigation bit, and we ignore Doppler and emission gain factors). The  direct electric field at the surface is  given by (refer to Figure~\ref{geometry})
\begin{equation}
E_I(t) = a\!\left[t-s/c\right] \, {e^{-i\omega t+in\pi} e^{ i  k s} \over 4\pi s }, 
\end{equation}
and the reflected electric field measured
at the receiver is given by 
\beq\label{realstarteq}
E(t)= {- i  \over 4 \pi} \int \sqrt{\cal G} \, {\cal R}\,  a\!\left[t-\left(r+s\right)/c\right]\,  e^{-i\omega t+in\pi}  { e^{ i  k(r+s)} \over r  s} \, (\vec{q} \cdot \hat{n})\,  dS , 
\eeq
where $\cal G$ is the antenna gain, $\omega$ is the carrier frequency (including Doppler offsets), $n\pi$ the (GPS) navigation bit,  $\cal R$ is the Fresnel coefficient, $k=2\pi/\lambda$, with $\lambda \approx$19 cm in GPS L1,  $r$ ($s$) is the distance between the receiver (transmitter) and each point of the sea-surface, $\hat{n}$ the  normal to the surface, and $\vec{q}=(\vec{q}_\bot, q_z)$ is  the scattering vector (the vector   normal to the plane that would specularly reflect the wave in the receiver direction). This vector is a function of the incoming and outgoing unit vectors $\hat{n}_i$ and $\hat{n}_s$, $\vec{q}= k(\hat{n}_i - \hat{n}_s)$. 

Equation~\ref{realstarteq} just states that each point in the surface acts like a local mirror source of the incoming field, incorporating the appropriate delay: the resulting field is the superposition of all those fields modulated by the antenna gain. Correlation of this signal with a replica of the ``clean'' direct signal will further modulate the support of this superposition, as we now discuss. \medskip

After correlation of the field signal\footnote{Which at this stage has been transformed into a current by the antenna, then again to a voltage, down-converted, digitized, etc., facts that we ignore here as they are not crucial to the discussion.} with an optimized code-carrier replica (such as $a[t-\tau]\exp(-i(\omega+\delta\omega)t]$) we obtain \cite{zavorotny2000},
\begin{equation} \label{starteq}
F(t)= {- i  e^{-i\Delta \omega t+in\pi} \over 4 \pi} \int {\cal M} {\cal R}\cdot { e^{ i  k(r+s)} \over r  s} \, (\vec{q} \cdot \hat{n})\,  dS ,
\end{equation} 
where $\Delta \omega$ is the residual carrier frequency.  The term ${\cal M}={\cal M}(\vec{\rho},z)=\sqrt{{\cal G}(\vec{\rho},z)}\, \chi(\vec{\rho},z)$ represents now here the antenna gain and Woodward Ambiguity Function (WAF) filtering on the surface. 
Here the WAF is given by 
\beq
\chi(\vec{\rho},t,\delta\tau, \delta \omega)= {1\over T_i}\int_t^{t+T_i} 
a(t') \,  a(t'+\delta\tau) e^{-i \delta \omega t'}
\eeq
and 
\beq
\delta\tau= \tau-(r+s)/c.
\eeq
In the present case, $\chi$ can be approximated by \cite{zavorotny2000}
\beq
\chi(\vec{\rho},t,\delta\tau, \delta f) =\Lambda(\delta \tau)\cdot S(\delta \omega)
\eeq
where
\bea
\Lambda(\delta \tau)&\equiv&{1\over T_i} \int_0^{T_i} a(t+t')\, a(t+t'+\delta\tau)dt' \nonumber \\
                 &=& 
\left \{ \begin{array}{ll}
1-{\delta\tau}/\tau_c, & |\delta\tau|  \leq \tau_c \\
-\tau/T_i, &  |\delta\tau|  >  \tau_c 
\end{array}
\right. .
\eea
and 
\beq
S(\delta \omega) \equiv {1\over T_i} \int_0^{T_i} e^{-i \delta \omega \, t' } dt' = 
{\sin (\delta\omega \, T_i/2) \over \delta \omega \, T_i/2 }  e^{-i \delta \omega \, T_i/2 }.
\eeq
Note that this function has support  near $\delta \omega \approx 0$, i.e., near $|\delta f| \leq 1/2T_i$.   
Similarly, $\Lambda (\delta \tau)$ provides most support to  the integration area satisfying $ |\tau- (r+s)/c)| \leq \tau_c$.  Thus, $F(t)$ effectively sums the field scattered from the surface at locations supported by the mentioned delay and Doppler conditions. The WAF's role is to simply modulate the support of the Fresnel integral.

The GNSS-R receiver measures the (real part) of the electric field (including the carrier) through  the coupling of the electric field to the antenna, which induces currents and voltage variations. This voltage is a direct measure of the electric field (polarization issues aside), up to a multiplicative constant, which we ignore here---and noise. GNSS signal processing (the correlation process) allows for removal of carrier and code modulation and thus the recovery of the electric field spatial and slow temporal  amplitude and phase variation as  filtered by the WAF (as is nicely described in \cite{zavorotny2000}). This is what Equation~\ref{starteq} represents. \medskip

We assume here that $\vec{q} \cdot \hat{n}\approx k$ (small slope approximation and a small patch approximation, with scattering and/or support only near the geometric flat surface specular point, or high-frequency limit).
We first  note that\footnote{As a convention, general vectors will be in {\bf bold} face, while vectors laying in the plane will be displayed with an arrow on to$\vec{\mbox{p}}$. A h$\hat{\mbox{a}}$t will always indicate a unit vector. We will also write, for any vector, $\mathbf{w}=\vec{w}_\bot +w_z \mathbf{\hat{z}} $.} (Figure~\ref{geometry})
\begin{equation}
\mathbf{ \mathbf{r'}}+\mathbf{s'}=-\mathbf{q}.
\end{equation}
We now write
$
\mathbf{s}=\mathbf{s'}+ \mathbf{\delta r}
$
and, for the emitter far field case which applies here (the emitting GPS satellite is very far in comparison with the scattering region size),
\begin{equation}
s=\sqrt{s'^2+(\delta r)^2+2\mathbf{s'}\cdot\mathbf{\delta r}} \approx s'\sqrt{1+ 2{\mathbf{\hat{s}'}\cdot \mathbf{\delta r} \over s'}} \approx s'+\mathbf{\hat{s}'}\cdot  \mathbf{\delta r}.
\end{equation}
Hence, we can rewrite Equation~\ref{starteq} as 
\begin{equation}
F(t)\approx {- i  k e^{-i\Delta \omega t+in\pi} \over 4 \pi} {e^{ i  k s'}\over s'} \int  {\cal M} {\cal R}\, { e^{ i  k(r +\mathbf{\hat{s}'} \cdot\mathbf{\delta r})} \over r } \,  dS .
\end{equation}
Now, the incoming field at the receiver   is given by 
\begin{equation}
I(t)\approx {e^{-i\Delta \omega t+in\pi} e^{ i  k|| \mathbf{s'} - \mathbf{r'}||} \over 4\pi s' }\approx 
 { e^{ i  ks'-k\mathbf{\hat{s}'}\cdot {\mathbf{r'}}} \over 4\pi s' },
\end{equation}
since $|| \mathbf{s'} - \mathbf{r'}||\approx s' -\mathbf{\hat{s}'}\cdot\mathbf{r'}$. We can now write the emitter far field expression for the ICF (F/I) as
\begin{equation}
F_I(t)\approx {- i  k } e^{i k\mathbf{\hat{s}'}\cdot \mathbf{r'}}  \int  {\cal M} {\cal R}\, { e^{ i  k(r +\mathbf{\hat{s}'} \cdot \mathbf{\delta r})} \over r  } \,  dS ,
\end{equation}
and writing $r'\approx r_o -H z/r_o$, we finally have
\begin{equation}
F_I(t)\approx {- i  k } e^{i k\mathbf{\hat{s}'}\cdot \mathbf{r'}}  \int  {\cal M} {\cal R}\cdot { e^{ i  k(r_o  -  H  z/r_o + \mathbf{\hat{s}'} \cdot\mathbf{\delta r} )} \over r  } \,  dS .
\end{equation}
The exponent has now been expanded to first order in $z$, and the Gaussian approximation for the correlation function can now be used. We can further approximate this result by writing $H/r_o\approx \sin\epsilon $ and $\mathbf{\hat{s}'} \cdot\mathbf{\delta r}\approx \mathbf{\hat{s}'}_\bot \cdot {\vec{\rho}} -\sin\epsilon$
\begin{equation}
F_I(t)\approx {- i  k } e^{i k\mathbf{\hat{s}'}\cdot \mathbf{r'}}  \int  {\cal M} {\cal R}\, { e^{ i  k(r_o  -  2z \sin\epsilon  + \mathbf{\hat{s}'}_\bot \cdot {\vec{\rho}} )} \over r  } \,  dS .
\end{equation}
We now expand the $r_o$ term in powers of $\Delta\vec{\rho}$,
\begin{eqnarray}
r_o&=&|| \mathbf{r'}+{\vec{\rho}}||^2 \\
&\approx& r'+\mathbf{\hat{r}'}_\bot \cdot {\vec{\rho}}+{1\over 2r'}(\mathbf{\hat{r}'}_\bot \cdot {\vec{\rho}})^2.\nonumber
\end{eqnarray}
Keeping only first order terms leads to the so-called Fraunhofer approximation. A step beyond the Fraunhofer approximation is obtained by keeping the second order term. We call this the Modified Fraunhofer approximation.

Keeping the second order term we can then write
\begin{equation}
F_I(t)\approx {- i  k } {e^{i k\mathbf{\hat{s}'}\cdot  \mathbf{r'}+ i  k r'} \over r'} \int  {\cal M} {\cal R}\, e^{i(   -  2k z \sin\epsilon  -\vec{q}_\bot \cdot {\vec{\rho}}+{k\over 2r'}(\mathbf{\hat{r}'}_\bot\cdot {\vec{\rho}})^2 )}  \,  dS .
\end{equation}
and finally ($ \mathbf{\hat{s}'}\cdot  \mathbf{r'}+ r'=\mathbf{r'}\cdot (\mathbf{\hat{s}'}+\mathbf{\hat{r}})=2iH\sin \epsilon$)
\begin{equation} \label{eq:MF}
F_I(t)\approx {- i  k } {e^{i2  kH\sin\epsilon} \over r'} \int  {\cal M} {\cal R}\, e^{i(   -  2kz\sin\epsilon  -\vec{q}_\bot \cdot {\vec{\rho}}+{k\over 2r'}(\mathbf{\hat{r}'}_\bot\cdot {\vec{\rho}})^2 )}  \,  dS .
\end{equation}


\section{Fraunhofer Approximation}

As an exercise, we carry out first the Fraunhofer approximation to the Fresnel integrand. We will see later that this approximation is not capable of representing directional dependence.

Dropping the second order term in Equation~\ref{eq:MF} we get
\begin{equation}
F_I(t)= - i  k {e^{ i  2kH\sin \epsilon} \over  r'} 
\int {\cal M} {\cal R}\, e^{- i \mathbf{q}\cdot \delta\mathbf{r}} \,  dS .
\end{equation}
Now we can compute the ICF autocorrelation function,
\begin{equation}
\Gamma(\Delta t)= \langle {-k^2 \over  r'^2} 
\int {\cal M} {\cal M}' {\cal R} {\cal R}' \, e^{- i \mathbf{q}\cdot \Delta( \delta\mathbf{r})} \,  dSdS'\rangle ,
\end{equation}
and writing $\delta\mathbf{r}=[\vec{\rho}, z]$, and $\mathbf{q}=[\vec{q}_\perp, {q}_z]$
\begin{equation}
\Gamma(\Delta t)= {-k^2 \over  r'^2} 
\int {\cal M} {\cal M}' {\cal R} {\cal R}' \, \langle e^{- i \vec{q}_\perp \cdot \Delta\vec{\rho} - {q}_z \Delta z}\rangle \,  dSdS' ,
\end{equation}
which (using $q_z=2k\sin\epsilon$) we can rewrite---assuming Gaussian statistics for the surface to write  \cite{beckmann} 
\begin{equation}
\langle e^{-2 i  k\sin\epsilon[z({\vec{\rho}},t)-z({\vec{\rho'}},t+\Delta t)]}\rangle_{z} = e^{ -4k^2\sin^2\! \epsilon\, \sigma_{z}^2[1-C(\Delta \vec{\rho},\Delta t)]},
\end{equation}
where $\sigma_{z}$ is the standard deviation of the surface elevation and $C(\Delta \vec{\rho},\Delta t)$ the spatio-temporal autocorrelation function of the surface---as
\begin{eqnarray}
\Gamma(\Delta t)&=& {-k^2 \over  r'^2} 
\int {\cal M} {\cal M}' {\cal R} {\cal R}' \,    e^{i\vec{q}_\perp \cdot \Delta{\vec{\rho}} -4k^2\sin^2\! \epsilon\, \sigma_{z}^2[1-C(\Delta {\vec{\rho}},\Delta t)]} \,  dSdS' .
\end{eqnarray}
We ignore, for the time being, WAF, antenna pattern and  Fresnel coefficients space dependence to write
\begin{equation}
\Gamma(\Delta t)\approx {{\cal A} {\cal R}^2 \over  r'^2} 
\int  e^{i\vec{q}_\perp \cdot \Delta{\vec{\rho}} -4k^2\sin^2\! \epsilon\, \sigma_{z}^2[1-C(\Delta \vec{\rho},\Delta t)]} \,  dSdS' .
\end{equation}
Ignoring these terms is licit at small $\Delta t$: then the integrand is strongly suppressed by the autocorrelation function itself. Nonetheless, in the next section we will keep them for completeness and clarity.

Now we write a simple Gaussian expression for the autocorrelation function and use the real version of the stationary phase approximation, the saddle-point approximation, to keep terms in $\Delta\vec{\rho}$ only to second order,
\begin{eqnarray}
C(\Delta \vec{\rho},\Delta t)& =& e^{-{\Delta t^2\over 2\tau_z^2} - \Delta \vec{\rho}  M \Delta \vec{\rho}}\nonumber \\
&\approx&  e^{-{\Delta t^2\over 2\tau_z^2}}\left(1 - \Delta \vec{\rho}M \Delta \vec{\rho}\right), \label{dircorr1}
\end{eqnarray}
where

\begin{eqnarray}
M &=& {1\over 2} R_{\psi}
\cdot
\left[
\begin{array}{cc} 1/l_u^2  & 0 \\ 0 & 1/l_c^2 \end{array}
\right]
\cdot
R_{-\psi} \\
&=& {1\over 2}
\left[
\begin{array}{cc} \cos \psi  & \sin \psi \\ -\sin \psi & \cos \psi \end{array}
\right]
\cdot
\left[
\begin{array}{cc} 1/l_u^2  & 0 \\ 0 & 1/l_c^2 \end{array}
\right]
\cdot
\left[
\begin{array}{cc} \cos \psi  & -\sin \psi \\ \sin \psi & \cos \psi \end{array}
\right] \nonumber
\end{eqnarray}
 and  where $l_u$ and $l_c$ are the up and cross ``wave'' correlation lengths, and $\psi$ the direction associated to $l_u$. Another way to write $M$ (using Dirac notation) is 
\begin{equation}
M={1\over 2 l_u^2}|{\hat v}\rangle\langle {\hat v} \mid +{1\over 2 l_c^2} \mid \hat{v}^\perp\rangle\langle {\hat{v}^\perp}\mid.
\end{equation}
This expression for the autocorrelation function is directional, but does not include a surface current---a key aspect to see directional effects in the ICF, as we will see later.
Now, define
\begin{equation}
\Xi[\Delta t]\equiv 4k^2\sin^2\! \epsilon\, \sigma_{z}^2 e^{-{\Delta t^2\over 2\tau_z^2}}. 
\end{equation}
Then we use
\begin{equation} \label{theintegral}
I=\int e^{-xAx+bx} d^nx = e^{-bA^{-1}b/4} \sqrt{\pi^n\over \det{A}},
\end{equation}
to calculate
\begin{eqnarray}
\Gamma(\Delta t)&\approx& \\
&\approx&  {A {\cal R}^2 e^{-4k^2\sin^2\! \epsilon\, \sigma_{z}^2 } \over  r'^2} 
\int  e^{i\vec{q}_\perp \cdot \Delta\vec{\rho} + 4k^2\sin^2\! \epsilon\, \sigma_{z}^2[C(\Delta \vec{\rho},\Delta t)]} \,  dSdS' \nonumber\\
 &=& 
{A {\cal R}^2 e^{-4k^2\sin^2\! \epsilon\, \sigma_{z}^2 } \over  r'^2} 
\int  e^{i\vec{q}_\perp \cdot \Delta{\rho} + \Xi[\Delta t]\left[ 1- \Delta \vec{\rho} M \Delta \vec{\rho}\right] } \,  dSdS' \nonumber \\
&=& 
{A {\cal R}^2 e^{-4k^2\sin^2\! \epsilon\, \sigma_{z}^2 +\Xi[\Delta t]} \over  r'^2} 
\int  e^{i\vec{q}_\perp \cdot \Delta{\rho} - \Xi[\Delta t] \Delta \vec{\rho} M \Delta \vec{\rho} } \,  dSdS' \nonumber \\
&=& 
{A {\cal R}^2 e^{-4k^2\sin^2\! \epsilon\, \sigma_{z}^2 +\Xi[\Delta t]\left (1 -\left(\vec{q}_\perp \right) M^{-1} \vec{q}_\perp /2\right)} \over  r'^2} \sqrt{\pi^2\over \det\left( \Xi[\Delta t]M^{-1}  \right)} \nonumber
 \\
&=& 
{A \pi {\cal R}^2 e^{-4k^2\sin^2\! \epsilon\, \sigma_{z}^2 +\Xi[\Delta t]\left (1 -\left(\vec{q}_\perp \right) M^{-1} \vec{q}_\perp /2\right)} \over  |\Xi[\Delta t]| r'^2} \sqrt{\det\left( M  \right)} \nonumber.
\end{eqnarray}

Now, many approximations have been used so far. The important thing to retain is that the autocorrelation function is suppressed by the SWH dependent term and that the coherence function can be written in the form,
\begin{equation}
\Gamma(\Delta t)\approx {\cal A}  e^{-4k^2\sin^2\! \epsilon\, \sigma_{z}^2}{ e^{ -\Xi[\Delta t] \left(1-\left(\vec{q}_\perp \right) M^{-1} \vec{q}_\perp /2\right)} }.  
\end{equation}
That is, suppressed by a SWH term, and with a temporal behavior modulated by the relationship of two directions: the surface autocorrelation direction, and the scattering axis.  We can again look at short times and write and updated version
of the autocorrelation function
 \begin{equation} \label{Corr_Func2new}
\Gamma(\Delta t) \approx A \, e^{-4k^2\sigma_{z}^2\frac{\Delta t^2}{2\tau_{z}^2}\sin^2\! \epsilon \left(1-\left(\vec{q}_\perp \right) M^{-1} \vec{q}_\perp /2\right)} .
\end{equation}
Since $\vec{q}_\perp=0$ in the near specular Fraunhofer  scenario, we do not have a directional modification. This is the expression used in \cite{soulat2004}. 
\medskip

Directional sensitivity is present here only in the case the instrument is pointing away from the specular direction. In the specular scenario (e.g., the antenna is pointed to the geometric specular point, or the antenna gain is very low), directional sensitivity of the ICF coherence time will  begin to appear only in the Modified Fraunhofer approximation {\em and} through the inclusion of a surface current term in the autocorrelation function, as we discuss next.

\section{Modified Fraunhofer}
A moment's  thought will show that the only way that a modulating term associated to the scattering axis can appear is through the higher order terms in the Fresnel integral---the Fraunhofer limit is not sufficient. The modified Fraunhofer provides us with the reference direction $\vec{r'}_\perp$ (the scattering axis).

We begin the discussion starting with Equation~\ref{eq:MF}:
$$
F_I(t)\approx {- i  k } {e^{2 i  kH\sin\epsilon} \over r'} \int  {\cal M} {\cal R}\, e^{i(   -  2k\sin\epsilon \,   z -\vec{q}_\bot \cdot {\vec{\rho}}+{k\over 2r'}(\mathbf{\hat{r}'}_\bot\cdot {\vec{\rho}})^2 )}  \,  dS .
$$
We will us a Gaussian term to simulate the impact of antenna gain and the WAF and write
\begin{equation}
F_I(t)\approx {- i  k \cal R} {e^{2 i  kH\sin\epsilon} \over r'}  \int   e^{i(   -  2k\sin\epsilon \,   z -\vec{q}_\bot \cdot {\vec{\rho}}+{k\over 2r'}(\mathbf{\hat{r}'}_\bot\cdot {\vec{\rho}})^2  +i \vec{\rho} R\vec{\rho}})  \,  dS ,
\end{equation}
where $R$ is a  regulating term to account for first chip zone, antenna gain, $\vec{q}\cdot \hat{n}$, etc. We will write 
\begin{equation}
O\equiv {k\over 2r'}\mid \mathbf{\hat{r}'}_\bot \rangle\langle \mathbf{\hat{r}'}_\bot |,
\end{equation}
so
\begin{equation}
N={k\over 2r'}\mid \mathbf{\hat{r}'}_\bot \rangle\langle \mathbf{\hat{r}'}_\bot | +i R = O +iR,
\end{equation}
and
\begin{equation}
F_I(t)\approx {- i  k \cal R} {e^{2 i  kH\sin\epsilon} \over r'}\int   e^{i(   -  2k\sin\epsilon \,   z -\vec{q}_\bot \cdot {\vec{\rho}}+\vec{\rho} N \vec{\rho}})  \,  dS .
\end{equation}

We can now rewrite
\begin{equation}
\Gamma(\Delta t)\approx {A {\cal R}^2 \over  r'^2} 
\int  e^{i\vec{q}_\perp \cdot \Delta{\vec{\rho}} -4k^2\sin^2\! \epsilon\, \sigma_{z}^2[1-C(\Delta \vec{\rho},\Delta t)] +i \vec{\rho} N \vec{\rho}- i \vec{\rho'} N^* \vec{\rho'} } \,  dSdS' .
\end{equation}

The next step is to define a wave direction. This is in fact defined already by the matrix $M$, but this directionality is not yet time-associated (which we need if we are to see the impact on the coherence time). We have to redefine the surface autocorrelation function to include the impact of something like a current.
Now we write,
\begin{eqnarray}
C(\Delta \vec{\rho},\Delta t)& =& \exp \left[  
-\left( \Delta \vec{\rho}-\vec{v}\Delta t  \right) 
M 
\left( \Delta \vec{\rho}-\vec{v}\Delta t    \right)   
\right]   \exp\left[ - A (\Delta t)^2  \right]  \\
&=& \exp \left[
-\Delta \vec{\rho} M \Delta \vec{\rho}
-(\Delta t)^2 {v^2 \over 2  l_u^2}
+\Delta t \vec{v}\cdot\Delta \vec{\rho} / l_u^2 
\right]\exp\left[ - A (\Delta t)^2  \right] \nonumber \\
&\approx&  \exp\left[ {-{\Delta t^2  {v^2 \over 2 l_u^2} }}\right] \left(1 - \Delta \vec{\rho} M \Delta \vec{\rho} +\Delta t \vec{v}\cdot\Delta \vec{\rho} / l_u^2  \right)\exp\left[ - A (\Delta t)^2  \right], \nonumber
\end{eqnarray}
where we have used $M\vec{v}=\vec{v} /(2l_u^2)$.

To be consistent with the earlier discussion, we  rewrite this as
\begin{eqnarray}\label{dircorr2}
C(\Delta \vec{\rho},\Delta t)& =&  e^{-{\Delta t^2   \over 2 \tau_z^2}}  \left(1 - \Delta \vec{\rho} M \Delta \vec{\rho} +\beta {\Delta t {\hat{v}}\cdot\Delta \vec{\rho} \over l_u \tau_z}  \right) \nonumber \\
& \equiv &  e^{-{\Delta t^2   \over 2 \tau_z^2}}  \left(1 -\Delta \vec{\rho} M \Delta \vec{\rho}   +\vec{p}\cdot \Delta{\rho} \right)
\end{eqnarray}
with $\vec{p}=\beta \Delta t \hat{v} /( l_u \tau_z)$. We expect $\beta<<1$. 
Comparing this equation with Equation~\ref{dircorr1} we see we now have a new term, the spatio-temporal cross term. This expression for the autocorrelation function says that if an observer moves along in the  ``current'' reference frame the surface will appear to decorrelate more slowly.

We now have
\begin{eqnarray}
\Gamma(\Delta t)&\approx& {A {\cal R}^2 \over  r'^2} e^{ -4k^2\sin^2\! \epsilon\, \sigma_{z}^2} 
\int  e^{i\vec{q}_\perp \cdot \Delta{\vec{\rho}} +4k^2\sin^2\! \epsilon\, C(\Delta \vec{\rho},\Delta t) +i \vec{\rho} N \vec{\rho}- i \vec{\rho'} N^* \vec{\rho'} } \,  dSdS' \nonumber\\
&\approx& {A {\cal R}^2 \over  r'^2} e^{ -4k^2\sin^2\! \epsilon\, \sigma_{z}^2} 
\int  e^{i\vec{q}_\perp \cdot \Delta{\vec{\rho}} +\Xi[\Delta t] \left(1 - \Delta \vec{\rho} M \Delta \vec{\rho}   +\vec{p}\cdot \Delta{\rho} \right) +i \vec{\rho} N \vec{\rho}- i \vec{\rho'} N^* \vec{\rho'} } \,  dSdS'\nonumber \\
&\approx& {A {\cal R}^2 \over  r'^2} e^{ -4k^2\sin^2\! \epsilon\, \sigma_{z}^2 + \Xi[\Delta t]} 
\int  e^{i(\vec{q}_\perp - i \Xi[\Delta t]\vec{p}) \cdot \Delta{\vec{\rho}} -\Xi[\Delta t] \left( \Delta \vec{\rho} M \Delta \vec{\rho}\right) +i \vec{\rho} N \vec{\rho}- i \vec{\rho'} N^* \vec{\rho'} } \,  dSdS'\nonumber
\end{eqnarray}
Define now
\begin{eqnarray}
\vec{\rho}_{_\oplus} &\equiv& \vec{\rho}+\vec{\rho'}, \nonumber\\
\vec{\rho}_{_\ominus} &\equiv& \vec{\rho}-\vec{\rho'}, \nonumber
\end{eqnarray}
hence
\begin{eqnarray}
\vec{\rho}  &=& {1\over 2} (\vec{\rho}_{_\oplus} + \vec{\rho}_{_\ominus}), \nonumber\\
\vec{\rho'} &=& {1\over 2} (\vec{\rho}_{_\oplus} -\vec{\rho}_{_\ominus}). \nonumber
\end{eqnarray}
Now, 
\begin{equation}
\vec{\rho} N \vec{\rho}-\vec{\rho'} N^* \vec{\rho'}  = {1\over 2}
\left\{\vec{\rho}_{_\oplus} (iR) \vec{\rho}_{_\oplus} + \vec{\rho}_{_\ominus} (iR) \vec{\rho}_{_\ominus} +
\vec{\rho}_{_\oplus} O \vec{\rho}_{_\ominus} \right\}. 
\end{equation}
We now have ($O=O^T$)
\begin{eqnarray}
\Gamma(\Delta t) &\approx& {A {\cal R}^2 \over  r'^2} e^{ -4k^2\sin^2\! \epsilon\, \sigma_{z}^2 + \Xi[\Delta t]} 
\int  e^{i(\vec{q}_\perp - i \Xi[\Delta t]\vec{p}+ {O\vec{\rho}_{_\oplus}\over 2}) \vec{\rho}_{_\ominus}
 - {1\over 2}  \vec{\rho}_{_\ominus}\left( 2\Xi[\Delta t]M+R\right) \vec{\rho}_{_\ominus} -{1\over 2}
\vec{\rho}_{_\oplus} R \vec{\rho}_{_\oplus}}\,  dS_{_\oplus} dS_{_\ominus}  .\nonumber 
\end{eqnarray}
We now use Equation~\ref{theintegral}. We compute first the ``+'' integral (with $b=iO\vec{\rho}_{_\ominus}/ 2$ and $A=R/2$), and then the ``--'' one,
\begin{eqnarray}
\Gamma(\Delta t) &\approx& {A {\cal R}^2 \over  r'^2} e^{ -4k^2\sin^2\! \epsilon\, \sigma_{z}^2 + \Xi[\Delta t]} 
\int  e^{i(\vec{q}_\perp - i \Xi[\Delta t]\vec{p}) \vec{\rho}_{_\ominus}
 - {1\over 2}  \vec{\rho}_{_\ominus}\left( 2\Xi[\Delta t]M+R+OR^{-1}O\right) \vec{\rho}_{_\ominus}} \,\sqrt{\pi^2\over \det\left(R/2\right)}\,   dS_{_\ominus}  \nonumber\\
 \Gamma(\Delta t) &\approx& {2\pi A {\cal R}^2 \over r'^2\sqrt{\det(R)} } e^{ -4k^2\sin^2\! \epsilon\, \sigma_{z}^2 + \Xi[\Delta t]} 
\int  e^{i(\vec{q}_\perp - i \Xi[\Delta t]\vec{p}) \vec{\rho}_{_\ominus}
 - {1\over 2}  \vec{\rho}_{_\ominus}\left( 2\Xi[\Delta t]M+R+OR^{-1}O\right) \vec{\rho}_{_\ominus}} \,   dS_{_\ominus},  \nonumber\\
\end{eqnarray}
and, finally
\begin{eqnarray}
\Gamma(\Delta t)&\approx& {2\pi A {\cal R}^2 \over r'^2\sqrt{\det(R)} } e^{ -4k^2\sin^2\! \epsilon\, \sigma_{z}^2 + \Xi[\Delta t]-{1\over 2} (\vec{q}_\perp - i \Xi[\Delta t]\vec{p})\cdot \left( 2\Xi[\Delta t]M+R+OR^{-1}O\right)^{-1}\cdot (\vec{q}_\perp - i \Xi[\Delta t]\vec{p})} \nonumber \\
&& \; \; \cdot  \sqrt{ 4 \pi^2\over \det\left( 2\Xi[\Delta t]M+R+OR^{-1}O\right) }  \nonumber\\
&=& {2\pi A {\cal R}^2 \over r'^2\sqrt{\det(R)} } e^{ -4k^2\sin^2\! \epsilon\, \sigma_{z}^2 + \Xi[\Delta t]-{1\over 2} (\vec{q}_\perp - i \Xi[\Delta t]\vec{p}) \cdot Q^{-1}\cdot  (\vec{q}_\perp - i \Xi[\Delta t]\vec{p})}  \,  \sqrt{ 4 \pi^2\over \det Q }   \nonumber\\
\end{eqnarray}
with $Q=2\Xi[\Delta t]M+R+OR^{-1}O$.
Finally, using $\vec{q}_\perp=0$ for the specular situation, we have
\begin{equation}
\Gamma(\Delta t)\approx {2\pi A {\cal R}^2 \over r'^2\sqrt{\det(R)} } e^{ -4k^2\sin^2\! \epsilon\, \sigma_{z}^2 + \Xi[\Delta t]+{1\over 2} \Xi[\Delta t]^2\vec{p}\cdot Q^{-1}\cdot \vec{p}}  \,  \sqrt{ 4 \pi^2\over \det Q }  \label{Qauto}
\end{equation}
\subsection{Isotropic antenna gain}
Recall that 
$$
O={k\over 2r'}\mid \mathbf{\hat{r}'}_\bot \rangle\langle \mathbf{\hat{r}'}_\bot |,
$$
and write
\begin{equation}
R= I/d^2,
\end{equation}
where $d$ is a length scale set by the antenna gain. Let us write
\begin{equation}
| \mathbf{\hat{r}'}_\bot \rangle = \cos \epsilon |\hat{u}_\phi\rangle.
\end{equation}
  We can now compute the determinant and inverse of the 2d matrix
\begin{equation}
{\cal Q} =2\Xi[\Delta t]M+R+OR^{-1}O.
\end{equation}
Since $\langle \mathbf{\hat{r}'}_\bot | \mathbf{\hat{r}'}_\bot \rangle= \cos^2 \epsilon$,
\begin{equation}
O^2=\left( k \cos^2 \epsilon \over 2r'\right)^2  \mid \hat{u}_\phi \rangle\langle\hat{u}_\phi |
\end{equation}
and, also using the definition of $M$,
\begin{eqnarray}
{\cal Q} &=&     
{\Xi[\Delta t] \over  l_u^2}|{\hat v}\rangle\langle {\hat v} \mid +{\Xi[\Delta t] \over  l_c^2} \mid \hat{v}^\perp\rangle\langle {\hat{v}^\perp}\mid
+{I\over d^2} +\left( d \, k \cos^2 \epsilon \over 2r'\right)^2  \mid \hat{u}_\phi \rangle\langle\hat{u}_\phi | \\ 
&=&
\left( {\Xi[\Delta t] \over  l_u^2} + {1\over d^2}\right)     \mid {\hat v}\rangle\langle {\hat v} \mid +
\left( {\Xi[\Delta t] \over  l_c^2} +  {1\over d^2}  \right)  \mid \hat{v}^\perp\rangle\langle {\hat{v}^\perp}\mid
+\left( d \, k \cos^2 \epsilon \over 2r'\right)^2  \mid \hat{u}_\phi \rangle\langle\hat{u}_\phi |. \nonumber
\end{eqnarray}
Now we expand $\mid \hat{u}_\phi \rangle = \mid {\hat v}\rangle \langle {\hat v} \mid \hat{u}_\phi \rangle 
+ \mid {\hat v}^\perp\rangle \langle {\hat v}^\perp \mid \hat{u}_\phi \rangle \equiv \cos \varphi\mid {\hat v}\rangle  + \sin \varphi \mid {\hat v}^\perp\rangle $ to write
\begin{equation}
\mid \hat{u}_\phi \rangle\langle\hat{u}_\phi | = 
\cos^2\varphi \mid {\hat v}\rangle\langle {\hat v} \mid + 
\sin^2 \varphi \mid {\hat v}^\perp\rangle\langle {\hat v}^\perp \mid + 
\cos\varphi \sin\varphi \left( \mid {\hat v} \rangle\langle {\hat v}^\perp \mid + \mid {\hat v}^\perp\rangle\langle {\hat v} \mid \right).
\end{equation}
We now define the scale
\begin{equation} 
l_\nu\equiv{ 2 r' \over d k \cos^2 \epsilon}
\end{equation}
In the case of interest (a coastal application using the GPS C/A code in non-grazing angles, $l_\nu << 1$.
Hence
\begin{eqnarray}
{\cal Q} &=&  
\left( {\Xi[\Delta t] \over  l_u^2} + {1\over d^2} +
{\cos^2\varphi\over l_\nu^2 }  \right)  \mid {\hat v}\rangle\langle {\hat v} \mid +
\left( {\Xi[\Delta t] \over  l_c^2} +  {1\over d^2} +
{\sin^2\varphi  \over l_\nu^2} \right) \mid \hat{v}^\perp\rangle\langle {\hat{v}^\perp}\mid \nonumber \\
&& \: \: +
{\cos\varphi \sin\varphi \over l_\nu^2}\left( \mid {\hat v} \rangle\langle {\hat v}^\perp \mid + \mid {\hat v}^\perp\rangle\langle {\hat v} \mid \right). \nonumber \\
\end{eqnarray}
Note that $\varphi$ is the angle between the scattering and wave direction.
It is straightforward to write the inverse of this matrix, using
\begin{equation}
\left[
\begin{array}{cc} a  & b \\ c & d \end{array}
\right]^{-1} = {1\over ad-cb}
\left[
\begin{array}{cc} d  & -b \\ -c & a \end{array}
\right],
\end{equation}
as
\begin{eqnarray}
{\det \cal Q}\cdot  {\cal Q}^{-1} &=&  
\left( {\Xi[\Delta t] \over  l_c^2} +  {1\over d^2} +{ \sin^2\varphi  \over l_\nu^2} \right)     \mid {\hat v}\rangle\langle {\hat v} \mid +
\left( {\Xi[\Delta t] \over  l_u^2} + {1\over d^2} +{   \cos^2\varphi\over l_\nu^2} \right)
  \mid \hat{v}^\perp\rangle\langle {\hat{v}^\perp}\mid \nonumber \\
&& \: \: -
{\cos\varphi \sin\varphi\over l_\nu^2} \left( \mid {\hat v} \rangle\langle {\hat v}^\perp \mid + \mid {\hat v}^\perp\rangle\langle {\hat v} \mid \right). \nonumber  \\
\end{eqnarray}
For large $d$ we ignore the middle term. 
\begin{eqnarray}
\det Q&\approx& {\Xi[\Delta t]^2\over l_u^2 l_v^2} +{\Xi[\Delta t]\over l_\nu^2} \left( {\sin^2 \varphi \over l_u^2}+{\cos^2\varphi \over l_c^2}\right) \nonumber\\
&\approx& {\Xi[\Delta t]\over l_\nu^2 l^2}.
\end{eqnarray}
In the last step we assume a small degree of spatial anisotropy in the ocean spectrum ($l=l_u\approx l_c$) and we use  $l_\nu << 1$.

Finally
\begin{eqnarray}
{\cal Q}^{-1} &=&  
\left(   l_\nu^2  +{l^2 \sin^2\varphi  \over \Xi[\Delta t] } \right)     \mid {\hat v}\rangle\langle {\hat v} \mid +
\left( l_\nu^2  +{ l^2  \cos^2\varphi\over \Xi[\Delta t] } \right)
  \mid \hat{v}^\perp\rangle\langle {\hat{v}^\perp}\mid \nonumber \\
&& \: \: -
{l^2 \cos\varphi \sin\varphi\over\Xi[\Delta t] } \left( \mid {\hat v} \rangle\langle {\hat v}^\perp \mid + \mid {\hat v}^\perp\rangle\langle {\hat v} \mid \right). \nonumber  \\
 &\approx &  
\left( {l^2 \sin^2\varphi  \over \Xi[\Delta t] } \right)     \mid {\hat v}\rangle\langle {\hat v} \mid +
\left( { l^2  \cos^2\varphi\over \Xi[\Delta t] } \right)
  \mid \hat{v}^\perp\rangle\langle {\hat{v}^\perp}\mid \nonumber \\
&& \: \: -
{l^2 \cos\varphi \sin\varphi\over\Xi[\Delta t] } \left( \mid {\hat v} \rangle\langle {\hat v}^\perp \mid + \mid {\hat v}^\perp\rangle\langle {\hat v} \mid \right). 
\end{eqnarray}


Finally, returning to Equation~\ref{Qauto}, we can write
\begin{eqnarray}
\Gamma(\Delta t)&\approx& {2\pi A {\cal R}^2 \over r'^2\sqrt{\det(R)} } 
\, e^{ -4k^2\sin^2\! \epsilon\, \sigma_{z}^2 + \Xi[\Delta t]+{\beta^2\over 2} \Xi[\Delta t]^2 {\Delta t  \over  \tau_z}\cdot {\sin^2\varphi \over\Xi[\Delta t] } \cdot {\Delta t  \over  \tau_z} }  \,  \sqrt{ 4 \pi^2\over \det Q } \nonumber \\
  &\approx& {2\pi A {\cal R}^2 d^2 l_\nu l \over r'^2 } e^{ -4k^2\sin^2\! \epsilon\, \sigma_{z}^2 + \Xi[\Delta t]\left( 1+{\beta^2\over 2}  {(\Delta t)^2  \over  \tau_z^2} {\sin^2\varphi}\right) }  \,  \sqrt{ 4 \pi^2\over \Xi[\Delta t]} \nonumber \nonumber \\
 &\approx& {2\pi A {\cal R}^2 d^2 l_\nu l \over r'^2 } \, 
 e^{ -4k^2\sin^2\! \epsilon\, \sigma_{z}^2 +4k^2\sin^2\! \epsilon\, \sigma_{z}^2 e^{-{\Delta t^2\over 2\tau_z^2}} \left( 1+{\beta^2\over 2}  {(\Delta t)^2  \over  \tau_z^2} {\sin^2\varphi}\right) }  \,  \sqrt{ 4 \pi^2\over 4k^2\sin^2\! \epsilon\, \sigma_{z}^2 e^{-{\Delta t^2\over 2\tau_z^2}}} \nonumber \nonumber \\
 &\approx& {4\pi^2 A {\cal R}^2 d^2 l_\nu l \over r'^2 2k \sin\! \epsilon\, \sigma_{z} } \, 
e^{ -4k^2\sin^2\! \epsilon\, \sigma_{z}^2 +4k^2\sin^2\! \epsilon\, \sigma_{z}^2 e^{-{\Delta t^2\over 2\tau_z^2}} \left( 1+{\beta^2\over 2}  {(\Delta t)^2  \over  \tau_z^2} {\sin^2\varphi}\right) }  \,  e^{\Delta t^2\over 4\tau_z^2} \nonumber \nonumber \\ 
&\approx& {4\pi^2 A {\cal R}^2 d^2 l_\nu l \over r'^2 2k \sin\! \epsilon\, \sigma_{z} } \, 
e^{ -4k^2\sin^2\! \epsilon\, \sigma_{z}^2\left[ 1- \left( 1-{\Delta t^2\over 2\tau_z^2}\right) \left( 1+{\beta^2\over 2}  {(\Delta t)^2  \over  \tau_z^2} {\sin^2\varphi}\right) \right]   } \nonumber \\
&\approx& {4\pi^2 A {\cal R}^2 d^2 l_\nu l \over r'^2 2k \sin\! \epsilon\, \sigma_{z} } \, 
e^{ -4k^2\sin^2\! \epsilon\, \sigma_{z}^2\left[ {\Delta t^2\over 2\tau_z^2} \left( 1 -{\beta^2\sin^2\varphi}\right) \right]   } 
\end{eqnarray}
Now we can see that the combination of the Modified Fraunhofer expression for the field, and the new autocorrelation function bring two new directions to play: the scattering axis, and the wave direction. To first order, and allowing for empirical adjustment to our approximations, the new expected result will be
 \begin{equation} \label{Corr_Func2final}
\Gamma(\Delta t) \approx A \, e^{-4k^2\sigma_{z}^2\frac{\Delta t^2}{2\tau_{z}^2}\sin^2\! \epsilon \left(1- \beta^2 \sin^2 \varphi \right)} .
\end{equation}
The coherence time of the ICF is now given by the width (second order moment) of this Gaussian function,
\begin{equation} \label{CT_field0}
\tau_{F} = \frac{ \tau_{z}}{2 k \sigma_{z} \sin \epsilon\sqrt{1- \beta^2 \sin^2 \varphi} } = \frac{\lambda}{\pi \sin \epsilon\sqrt{1- \beta^2 \sin^2 \varphi }}\frac{\tau_{z}}{\mbox{SWH}}.
\end{equation}
According to this model, $\tau_{F}$  depends on the electromagnetic wavelength  and the ratio between the correlation time of the surface and the significant wave height (an inverse velocity). It is also apparent that the ICF is more coherent when the scattering direction and wave direction are perpendicular.


Equation \ref{CT_field0} should provide the basis to determine direction in the waves (Oceanpal SWH Algorithm 2). Using data from visible satellites at any given time (with different elevations and azimuths) the optimal wave direction $\phi_u$ and  ocean ``z-velocity'' $Z_v= \mbox{SWH}/\tau_z $ parameters can be searched for (as described, e.g., in \cite{soulat2004}). We first write $\varphi=\phi-\phi_v$, i.e., as the difference between the satellite azimuth and the wave direction. The we write, for each satellite in view,
\beq
\tau_{F}^k =\frac{\lambda}{\pi \sin \epsilon ^k \, \sqrt{1- \beta^2 \sin^2 (\phi^k-\phi_u) }}  {1\over Z_v}. 
\eeq
To make the measurement more robust, several Oceanpal instruments can be deployed in the area of interest. We then write and equation for each satellite-receiver link (Oceanpal SWH Algorithm 2):
\beq \label{final}
\tau_{F}^{i,k} =\frac{\lambda}{\pi \sin \epsilon ^{i,k} \, \sqrt{1- \beta^2 \sin^2 (\phi^{i,k}-\phi_u) }}{1\over Z_v}. 
\eeq
Finally, recall that  in \cite{soulat2004},  based  on the  {\it Elfouhaily et al.}  spectrum \cite{elfouhaily1997} we  derived a linear relationship between $\tau_z$ and the SWH: $\tau_z=a_s+b_s\ast \mbox{SWH}$ (with $a_s$=0.167, $b_s$=0.388, and an error of 0.03 s). This relation turns out to be rather independent of wave age. Using it, we can now rewrite Equation~\ref{final} as
\beq
\tau_{F}^{i,k} =\frac{\lambda}{\pi \sin \epsilon ^{i,k} \, \sqrt{1- \beta^2 \sin^2 (\phi^{i,k}-\phi_u) }}\frac{ a_s+b_s\ast \mbox{SWH}   }{\mbox{SWH}}.
\eeq

\section{Conclusion}
In this note we have derived an expression which can be used for the definition of a semi-empirical algorithms for low altitude, static GNSS-R sea state and wave direction (as in \cite{soulat2004}). The expression relates the  interferometric field autocorrelation function coherence time with sea state parameters.  Although directional sensitivity is already present in the Fraunhofer regime analysis with a moderate gain antenna factor pointing away from the specular, surface current sensitivity appears only in the Modified Fraunhofer regime. The derived expression is sensitive to the wave direction relative to the scattering direction and to the ocean ``z-velocity'', the ratio of SWH to surface coherence time, and can be used for coastal sea state monitoring using one or more receivers.

\section*{Acknowledgements}

This study was carried out under the Starlab Oceanpal Project. http://oceanpal.com. We thank Bertrand Chapron (IFREMER) for very useful discussions. 
 {\em All Starlab authors have contributed significantly; the Starlab author list has been ordered randomly}.

\end{document}